\documentclass[11pt]{article}
\usepackage{moriond,epsfig}

\def\be{\begin{equation}}
\def\ee{\end{equation}}
\def\bea{\begin{eqnarray}}
\def\eea{\end{eqnarray}}

\begin{document}
\vspace*{4cm}
\title{Searches for New Physics at D\O}

\author{ G. Brooijmans \\ for the D\O\ Collaboration }

\address{Fermi National Accelerator Laboratory, P.O. Box 500, \\
Batavia, IL 60510, USA.}

\maketitle\abstracts{The integrated luminosity at Run 2 of the Tevatron is
approaching the Run 1 total, and data analysis is progressing.  New 
results in searches for new physics by the D\O\ experiment are presented
in a variety of channels, demonstrating good performance of the 
detector and detailed understanding of the data.}

\section{Introduction}
\label{sec:intro}

The D\O\ experiment has been taking data at Run 2 of the Tevatron since
March 2001.  In the early part of the run, the data was used for detector 
commissioning and calibration, and only data taken since April 2002,
when the fiber tracker instrumentation was completed, is used for physics
analysis.  Between April 2002 and January 2003, the Tevatron delivered 
135 $pb^{-1}$ of integrated luminosity, of which 83 $pb^{-1}$ were 
recorded by D\O .  
It should be noted that D\O 's datataking efficiency 
increased gradually during this period, and reached 90\% at the end of 2002.
The results presented here use up to 50 $pb^{-1}$ from that data sample.

\section{Searches for Supersymmetry}
\label{sec:susy}

Signals for supersymmetry are searched for in a variety of signatures 
predicted by various models of supersymmetry breaking.  A few of these 
are presented here.
For a given superpartner mass, production cross-sections at the Tevatron are 
highest for colored particles and lowest for particles that only interact weakly.
However, while the cross-sections are therefore highest 
for squarks and gluinos, these are expected
to be significantly heavier than the electroweak gauginos, such that for a given 
point in parameter space the production rates are similar~\cite{sugrareport}.  

\subsection{Jets and Missing $E_T$}
\label{subsec:jetsmet}

The generic signature for production of squarks and/or gluinos in SUGRA-type
models consists of jets and missing transverse energy (missing $E_T$).  These
come from the cascade decays of the squarks and gluinos into quarks, gluons
and the Lightest Supersymmetric Particle (LSP).  In these models the latter is 
heavy (${\cal O} (100\ GeV)$),
neutral, stable and weakly interacting, such that it escapes the detector
undetected, leaving as its only signature an imbalance in transverse momentum.

\begin{figure}
\centerline{
\psfig{file=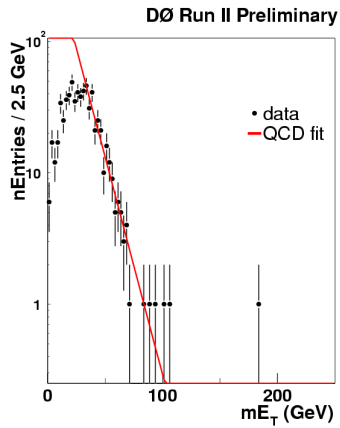,height=6cm}}
\caption{\label{fig:jetsmet} Distribution of jets + missing $E_T$ events as a function 
of missing $E_T$.  The data (points) is fitted (line) at moderate missing $E_T$ to 
estimate the QCD contribution to the high missing $E_T$ region.}
\end{figure}

This type of signature unfortunately suffers from very significant instrumental
backgrounds, due to the tails of the dominant standard model production of jets
(denoted QCD in the following).  A pilot analysis has been performed using 
4 $pb^{-1}$ of data in order
to demonstrate that these backgrounds could be understood and reduced 
sufficiently.  The analysis starts by applying a series of ``cleaning'' cuts
designed to reduce the instrumental background, followed by topological
cuts to increase the signal over background ratio.  After that, the physics
backgrounds are evaluated using simulation, and the QCD background is derived
from the data at moderate missing $E_T$.  This is illustrated in 
Figure~\ref{fig:jetsmet}.  As can be seen 
from Table~\ref{tab:jetsmet}, the number of 
\begin{table}
\caption{\label{tab:jetsmet} Expected and observed events for various cuts on the 
minimal event missing $E_T$.  The last line gives a limit on the new physics 
production cross-section of events with at least two hard jets (of which one 
has $p_T > 100\ GeV$) and large missing $E_T$.}
\vspace{0.4cm}
\centerline{
\begin{tabular}{|l|cccc|}
\hline
Cut: Missing $E_T >$ & 70 GeV & 80 GeV & 90 GeV & 100 GeV \\
\hline
\hline
Expected Events & 18.4 $\pm$ 8.4 & 9.5 $\pm$ 5.3 & 5.1 $\pm$ 3.2 & 2.7 $\pm$ 1.8 \\
\hline
Observed Events & 7 & 6 & 4 & 3 \\
\hline
95\% C.L. Cross-Section Limit (pb) & 4.2 & 3.8 & 3.1 & 2.7 \\
\hline
\end{tabular}}
\end{table}
data events at high missing $E_T$ is in good agreement with the expectations
and does not exhibit any anomalous tails.

\subsection{Di- and Trileptons}
\label{subsec:trileptons}

The cascade decays of the electroweak gauginos, which are predicted to be relatively
light, often lead to signatures with multiple leptons.  At hadron colliders, this 
is the SUGRA channel with the lowest background, and is often called the ``golden
channel''.  It should be noted however that good sensitivity to this signal 
requires both low lepton $p_T$ thresholds (because of the small mass difference
between the gauginos and the LSP), and the ability to identify hadronic tau decays.
This is due to the fact that at large $\tan \beta$ the leptonic branching fractions
for charginos and neutralinos are dominated by taus.  

\begin{figure}
\centerline{
\psfig{file=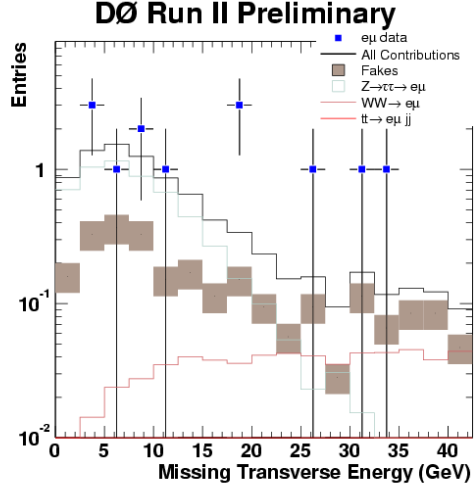,height=7cm}}
\caption{\label{fig:emux} Distribution of expected (histogram) and observed (points)
$e \mu$ events as a function of missing $E_T$.}
\end{figure}

Among the multilepton signatures, the channel with an electron and a muon has such
low backgrounds that the analysis is pursued in a model-independent way.  Events 
are required to have at least one electron and one muon, both with $p_T > 15\ GeV$.
The fake rates are estimated from data, while the physics backgrounds are evaluated
from simulation.  The dominant physics backgrounds are $Z \rightarrow \tau \tau$ at
low missing $E_T$, and $WW$ and $t \overline{t}$ production at high missing $E_T$.
In approximately 30 $pb^{-1}$ of data (the uncertainty on the luminosity is currently
conservatively estimated at 10\%), the data agree well with the backgrounds as can be 
seen from Figure~\ref{fig:emux}.
Based on this,
a model-independent upper limit on acceptance $\times$ cross-section is set for 
new physics leading to events with an electron and a muon as a function of missing
$E_T$.  This limit ranges from about 400 to 100 $fb$ for missing $E_T$ going from
0 to 35 $GeV$.

\begin{figure}
\centerline{
\psfig{file=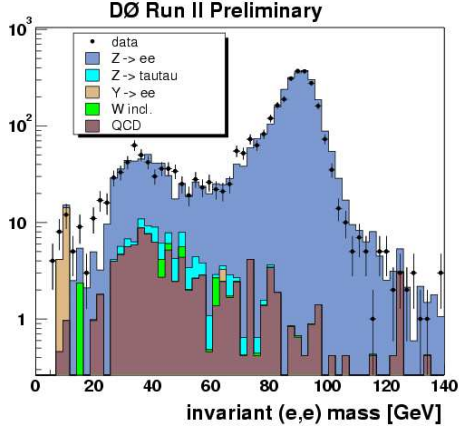,height=6cm}}
\caption{\label{fig:diem} Distribution of expected (histogram) and observed (points) 
$e e$ events as a function of dielectron invariant mass.}
\end{figure}

The study of channels with two electrons and a third lepton starts with the study
of the dielectron invariant mass spectrum.  This is shown in Figure~\ref{fig:diem}
and is used to check that trigger, reconstruction, simulation and the QCD fake
background are well understood.  While not perfect yet, the agreement is good.
This analysis uses approximately 40 $pb^{-1}$ of data, and the successive cuts,
along with the expected and observed numbers of events are given in 
Table~\ref{tab:diem}.  
\begin{table}
\caption{\label{tab:diem} Expected and observed events for successive cuts
on dielectron events.}
\vspace{0.4cm}
\centerline{
\begin{tabular}{|l|c|c|}
\hline
 & Backgrounds & Data \\
\hline
\hline
$p_T(e_1) > 15\ GeV, p_T(e_2) > 10\ GeV$ & 3216 $\pm$ 43.2 & 3132 \\
\hline
$10\ GeV < M_{ee} < 70\ GeV$ & 660.2 $\pm$ 19.1 & 721 \\
\hline
$M_T > 15\ GeV$ & 96.4 $\pm$ 8.1 & 123 \\
\hline
Additional Isolated Track, $p_T > 5\ GeV$ & 3.2 $\pm$ 2.3 & 3 \\
\hline
Missing $E_T > 15\ GeV$ & 0.0 $\pm$ 2.0 & 0 \\
\hline
\end{tabular}}
\end{table}
The typical selection efficiency for SUGRA events at points
in parameter space close to the current exclusion limit is 2-4\%, but the 
sensitivity is still about a factor of 7 away from extending the excluded area.  
In addition to increased integrated luminosity, efforts are underway to 
improve the sensitivity by improving the efficiencies and adding other channels.

\begin{figure}
\centerline{
\rotatebox{-90}{\psfig{file=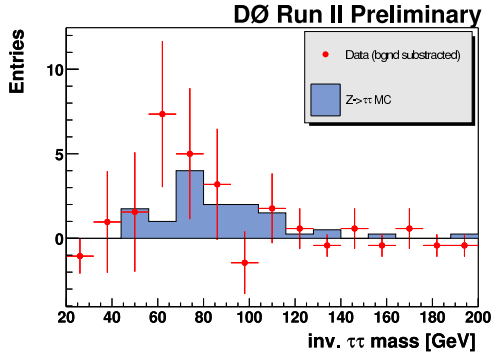,height=7cm}}}
\caption{\label{fig:etau} Distribution of candidate $Z \rightarrow \tau \tau X
\rightarrow e h X$ events (points) as a function of reconstructed ditau invariant
mass. Also shown is the expectation from simulation (histogram).}
\end{figure}

One of these approaches involves identifying $Z \rightarrow \tau \tau$ events where
one of the taus decays hadronically.  For the $p \overline{p} \rightarrow
Z X \rightarrow \tau \tau X \rightarrow
e h X$ channel, events with an electron with $p_T > 12\ GeV$ and a tau-like jet 
are selected from about 50 $pb^{-1}$ of data.  
The tau-like requirements are a cut on the jet width, and only 
one track with $p_T > 1.5\ GeV$ compatible with the tau mass (i.e. starting from
the leading track, the two-track invariant mass is calculated for all 
tracks with $p_T > 1.5\ GeV$ and required to be below the tau mass).  For each
of the tau decay channels (with or without a neutral pion), a neural
net using both calorimetric and tracking variables for inputs is used to discriminate
between QCD and tau jets.  After cutting on the neural net outputs, the ditau 
invariant mass is reconstructed under the hypothesis that the tau direction is the 
same as the visible tau daughter direction.  Finally, same-sign $e \tau$ events 
are subtracted from opposite sign events, leading to the histogram shown in
Figure~\ref{fig:etau}, 
indicating a clear signal which can be used to further 
improve the identification of hadronically decaying taus.

\subsection{Diphotons and Missing $E_T$}
\label{subsec:ggmet}

\begin{figure}
\centerline{
\psfig{file=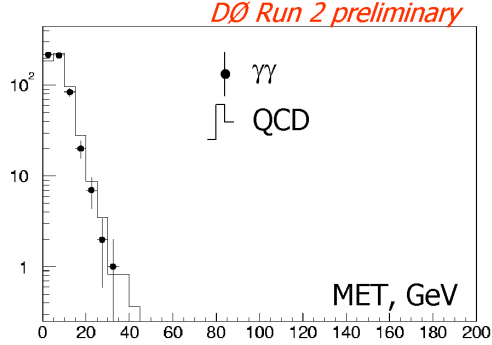,height=5cm}}
\caption{\label{fig:ggmet} Distribution of diphoton events (points) as a function of
missing $E_T$.  The histogram is the expected background from QCD fakes.}
\end{figure}

As opposed to the SUGRA scenario, in Gauge Mediated Supersymmetry Breaking (GMSB)
the LSP is a very light ($\ll eV$) gravitino, and the model phenomenology is driven
by the nature of the NLSP (Next-to-Lightest Supersymmetric Particle).  In the case
of a ``bino-like'' NLSP, events will have 2 energetic photons and large 
missing $E_T$.  This search uses about 50 $pb^{-1}$ of data, requires two photons
with $p_T > 20\ GeV$, and applies quality and topological cuts to reduce the 
instrumental backgrounds and increase the signal over background ratio.  The remaining
background is dominated by QCD fakes and is determined from data.  The resulting
missing $E_T$ distribution is shown in Figure~\ref{fig:ggmet}
and shows good agreement between data and backgrounds.  A limit
is set on GMSB using the Snowmass slope~\cite{snowmass} 
with $\Lambda$ as the only free parameter
and $M = 2\Lambda , N_5 = 1, tan \beta = 15$ and $\mu > 0$.  The limit corresponds
to a lower limit on the neutralino mass of $66\ GeV$ at 95\% C.L., very close to
the Run 1 result.

\section{Exotics}
\label{sec:exotics}

One of the more exotic signals for new physics D\O\ searches for is leptoquarks.
The analysis presented here uses about 30 $pb^{-1}$ of data to search for second
generation leptoquarks decaying to a muon and a $c$ or $s$ quark.  Events are 
required to have two opposite sign muons with $p_T > 15\ GeV$, two jets with 
$p_T > 20\ GeV$ and $M_{\mu \mu} > 110\ GeV$.  The dominant background comes
from $\gamma^* / Z \rightarrow \mu \mu + jets$, and since no events are found a lower 
limit on the leptoquark mass is set at $157\ GeV$, $43\ GeV$ worse than the Run 1
limit.

\section{Large Extra Dimensions}
\label{sec:led}

\begin{figure}
\centerline{
\psfig{file=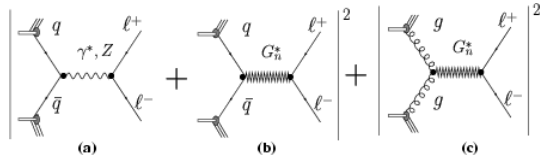,height=2cm}}
\caption{\label{fig:led1} Feynman diagrams showing contributions of virtual graviton 
exchange to Drell-Yan processes.}
\end{figure}

\begin{figure}
\centerline{
\psfig{file=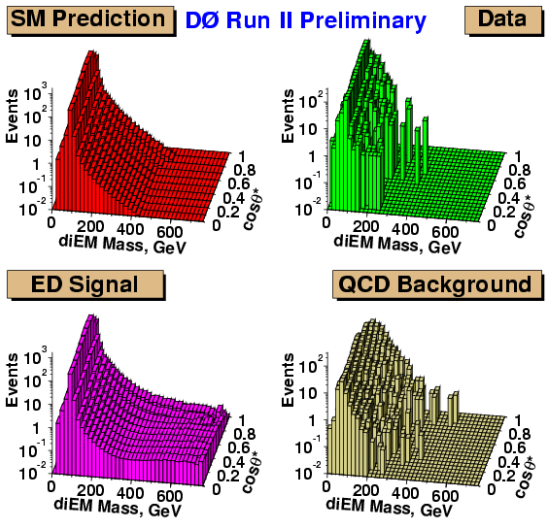,height=7cm}}
\caption{\label{fig:led} Standard model expectation, data, extra dimension signal and
QCD fake background in the invariant mass - $cos\theta^*$ plane for the di-em channel.}
\end{figure}

D\O\ searches for Large Extra Dimensions (LED) in the ADD~\cite{add} model
assuming standard model particles are confined to a 3-brane, but gravity 
propagates in the extra dimensions.  In these analyses, the signature is an 
excess of high-mass dielectron, diphoton or dimuon events 
due to the coupling to Kaluza-Klein gravitons.  Figure~\ref{fig:led1}
shows Feynman diagrams of the standard model and graviton contributions to
the dilepton processes, which can be parametrized as follows:
\begin{equation}
\frac{d^2 \sigma}{dM d\cos\theta^*} = f_{SM} + f_{interf} \eta_G +
                                     f_{KK} \eta_G^2 .
\end{equation}
Here $\cos\theta^*$ is the scattering angle in the di-em or dimuon rest frame, and
$\eta_G$ measures the contribution from gravitons.  The dielectron and diphoton 
channels are treated simultaneously (I.e. no track requirement is imposed) so that 
events are required to have 2 electromagnetic objects with $p_T > 25\ GeV$, and missing
$E_T < 25\ GeV$ to ensure the events are well measured.  In the dimuon channel, two
muons with $p_T > 15\ GeV$ and $M_{\mu \mu} > 40\ GeV$ are required.  In both cases,
physics backgrounds are derived from simulation and instrumental backgrounds from
data.  The data distribution in the invariant mass - $\cos\theta^*$ plane is then fitted 
to signal + background (see Figure~\ref{fig:led}) to evaluate an upper limit on $\eta_G$.  
This limit is translated into a lower limit on $M_S$, the fundamental Planck scale, with 
results shown in Table~\ref{tab:led} for various formalisms~\cite{led}.  
The dielectron and diphoton channel yields a 
\begin{table}
\caption{\label{tab:led} Lower limits in TeV on $M_S$, the fundamental Planck scale,
for various formalisms (see text).}
\vspace{0.4cm}
\centerline{
\begin{tabular}{|l|cccc|}
\hline
Formalism & GRW & HLZ, n=2 & HLZ, n=7 & Hewett, $\lambda = +1$ \\
\hline
\hline
Di-Em ($\approx 50 pb^{-1}$) & 1.12 & 1.16 & 0.89 & 1 \\
\hline
Dimuon ($\approx 30 pb^{-1}$) & 0.79 & 0.68 & 0.63 & 0.71 \\
\hline
\end{tabular}}
\end{table}
limit which is very close to the Run 1 result, while the dimuon channel is new at hadron
colliders.  Both results are competitive with LEP.

\section{Conclusions}

D\O\ continues to search for many different signatures for new physics.  The effects of 
increased center-of-mass energy and an improved detector can now be seen in improved 
sensitivity compared to Run 1, and efforts continue on calibration and reconstruction
to further extend the experiment's reach.  We are now entering uncharted territory,
so signal could be just around the corner.



\end{document}